\documentstyle[12pt]{article}
\newcommand{\bce}{\begin{center}}
\newcommand{\ece}{\end{center}}
\newcommand{\beq}{\begin{equation}}
\newcommand{\eeq}{\end{equation}}
\newcommand{\bea}{\vspace{0.25cm}\begin{eqnarray}}
\newcommand{\eea}{\end{eqnarray}}

\newcommand{\ba}{\begin{array}}
\newcommand{\ea}{\end{array}}

\newcommand{\doublespace}{
    \renewcommand{\baselinestretch}{1.6}\large\normalsize}

\def\lsim{\mathrel{\rlap{\lower4pt\hbox{\hskip1pt$\sim$}}
    \raise1pt\hbox{$<$}}}	  
\def\gsim{\mathrel{\rlap{\lower4pt\hbox{\hskip1pt$\sim$}}
    \raise1pt\hbox{$>$}}}	  

\def\Pom{{\bf I\!P}}

\def\lsim{\mathrel{\rlap{\lower4pt\hbox{\hskip1pt$\sim$}}
    \raise1pt\hbox{$<$}}}         
\def\gsim{\mathrel{\rlap{\lower4pt\hbox{\hskip1pt$\sim$}}
    \raise1pt\hbox{$>$}}}         

\def\Pom{{\bf I\!P}}

  \advance\oddsidemargin  by -1in
  \advance\evensidemargin by -1in
  \advance\topmargin      by -0.5in
\def\lsim{\mathrel{\rlap{\lower4pt\hbox{\hskip1pt$\sim$}}
    \raise1pt\hbox{$<$}}}         
\def\gsim{\mathrel{\rlap{\lower4pt\hbox{\hskip1pt$\sim$}}
    \raise1pt\hbox{$>$}}}         

\def\Pom{{\bf I\!P}}
\def\beq{\begin{equation}}
\def\endeq{\end{equation}}
\def\arr{\begin{eqnarray}}
\def\endarr{\end{eqnarray}}
\makeindex

\doublespace

\begin{document}


\phantom{.}{\bf \Large \hspace{10.0cm} KFA-IKP(Th)-1993-34 \\
\phantom{.}\hspace{11.7cm}26 November 1993\vspace{0.4cm}\\ }

\begin{center}
{\bf\sl \Large The $s$-channel approach to Lipatov's pomeron \\
and
hadronic cross sections}
\vspace{1.0cm}\\
{\bf \large
N.N.~Nikolaev$^{a,b}$, B.G.~Zakharov$^{a,b}$ and V.R.Zoller$^{a,c}$
\bigskip\\}
{\it
$^{a}$IKP(Theorie), KFA J{\"u}lich, 5170 J{\"u}lich, Germany
\medskip\\
$^{b}$L. D. Landau Institute for Theoretical Physics, GSP-1,
117940, \\
ul. Kosygina 2, Moscow 117334, Russia.\medskip\\
$^{c}$ Institute for Theoretical and Experimental Physics,\\
Bolshaya Cheremushkinskaya 25, 117259 Moscow, Russia.
\vspace{1.0cm}\\ }
{\Large
Abstract}\\
\end{center}

We derive a generalized Balitskii-Fadin-Kuraev-Lipatov
equation, which applies directly to the perturbative QCD component of
total cross section. With the
gluon correlation radius $R_{c} \sim 0.4$fm we reproduce
the empirical rate of growth of the hadron-nucleon total
cross sections. The simultaneous estimate of the triple-pomeron
coupling also agrees with the experiment.
 \bigskip\\

\begin{center}
    \large To be published in:{\sl~~ JETP Letters\bigskip\\ }
E-mail: kph154@zam001.zam.kfa-juelich.de
\end{center}

\pagebreak






Lipatov and his collaborators have shown [1-3] that
for short-distance interactions the QCD pomeron has the intercept
\beq
\alpha_{\Pom} = 1+\Delta_{\Pom} =
1 +  {12\log{2}\over \pi} \alpha_{S}  \, ,
\label{eq:1}
\endeq
which gives a very large $\Delta_{\Pom} \sim 1$ even with the reasonably
small strong coupling $\alpha_{S}=g_{S}^{2}/4\pi \sim 0.4$
appropriate for the already short distances  ($r\sim 0.15$fm) (here $g_{S}$
is the color charge). On the other hand, the
$\sigma_{tot} \propto s^{\Delta_{\Pom}(hN)}$ fit
of the hadronic total cross sections yields
$\Delta_{\Pom}(hN)\sim 0.1$ ~[4] (here $s$ is the square of the c.m.s.
energy).

In this paper we derive a particularly simple generalization of
the Balitskii-Fadin-Kuraev-Lipatov
(BFKL) equation directly for the total cross sections.
We use the
$s$-channel approach to the pomeron, which is based on the technique
of light-cone multiparton wave functions introduced
by the two of the present authors [5,6]. It allows
us to easily introduce in the gauge-invariant manner
the effect of a finite radius for gluon
correlations $R_{c}$ and to evaluate the effective intercept of
the pomeron for the hadronic scattering regime.
We relate the growth of
the total cross section to the rise of the number of perturbative
gluons in the lightcone hadrons.
The lightcone wave functions of the multiparton
states and their interaction cross sections
were derived in our previous paper [6] and applied to an
analysis of the diffractive deep inelastic scattering in the
Double-Leading-Logarithm Approximation (DLLA)
(see also [7]). In this paper we extend the considerations of [6] to the
BFKL regime, discuss the connection between the BFKL and DLLA
regimes, and estimate $\Delta_{\Pom}$ for hadronic scattering processes.




Our starting point is the lowest order perturbative
QCD cross section
for the scattering of the two color dipoles
$\vec{r}$ and $\vec{R}$ (here $\vec{r},\vec{R}$ are
2-dimensional vectors in the impact parameter plane)
\beq
\sigma_{0}(\vec{r},\vec{R})=
{32 \over 9}
\int {d^{2}\vec{k}
\over(k^{2}+\mu_{G}^{2})^{2} }
\alpha_{S}^{2}
\left[1-\exp(-i\vec{k}\vec{r})\right]
\left[1-\exp(i\vec{k}\vec{R})\right]  \, .
\label{eq:2}
\endeq
Here the effective mass of the gluon $\mu_{G}$ serves as a
reminder that the colour forces can not propagate beyond the
the gluon correlation radius
$R_{c}= 1/\mu_{G}$ and $\alpha_{S}^{2}$ must be understood
as $\alpha_{S}({\rm max}\{k^{2},{1\over r^{2}}\})
\alpha_{S}({\rm max}\{k^{2},{1\over R^{2}}\})$ [5].
At $r \ll R \lsim R_{c}$, Eq.~(\ref{eq:2}) gives the driving
term of the DLLA cross section (for a detailed discussion of
the DLLA regime see [6])
\beq
\sigma_{0}(r,R)\approx Cr^{2}
\alpha_{S}(r)L(R,r)\,\, ,
\label{eq:3}
\endeq
which is independent of $R_{c}$. Here
$L(R,r)\approx\log[\alpha_{S}(R)/\alpha_{S}(r)]$.
In terms of the dipole-dipole cross section (\ref{eq:2}) the
perturbative part of the total cross section for the interaction
of mesons $A$ and $B$ equals
\arr
\sigma^{(pt)}(AB)=\langle \langle \sigma(r_{A},r_{B}) \rangle_{A}
\rangle_{B}= ~~~~~~~~~~~~~~~~~~\nonumber\\
\int dz_{A} d^{2}\vec{r}_{A}
dz_{B} d^{2}\vec{r}_{B}
|\Psi(z_{A},\vec{r}_{A})|^{2}
|\Psi(z_{B},\vec{r}_{B})|^{2}
\sigma_{0}(\vec{r}_{A},\vec{r}_{B}) \,.
\label{eq:4}
\endarr
The advantage of the representation
(\ref{eq:4}) is that it makes full use of the
exact diagonalization of the scattering matrix
in the dipole-size representation. Hereafter we discuss
$\sigma(\vec{r},\vec{R})$  averaged over the relative
orientation of dipoles.

The perturbative $q\bar{q}g$ Fock state generated radiatively
from the parent colour-singlet $q\bar{q}$ state
has the interaction cross section $
\sigma_{3}(r,\rho_{1},\rho_{2})
=
{9 \over 8}[\sigma_{0}(\rho_{1})+
\sigma_{0}(\rho_{2})]-{1\over 8}\sigma_{0}(r)$,
where $\vec{\rho}_{1,2}$ are separations of the
gluon from the quark and antiquark respectively,
$\vec{\rho}_{2}=\vec{\rho}_{1}+\vec{r}$ [6].
The increase of the cross section for the presence of
gluons equals  (we suppress the target variable $R$)
\beq
\Delta \sigma_{g}(r,\rho_{1},\rho_{2})=\sigma_{3}(r,\rho_{1},\rho_{2}) -
\sigma_{0}(r)=
{9 \over 8}[\sigma_{0}(\rho_{1})+
\sigma_{0}(\rho_{2})-\sigma_{0}(r)]   \, \, ,
\label{eq:5}
\endeq
The lightcone density of soft, $z_{g}\ll 1$,
 gluons in the $q\bar{q}g$ state derived
in [6] equals
\arr
|\Phi_{1}(\vec{r},\vec{\rho}_{1} ,\vec{\rho}_{2},z_{g})|^{2}=
{}~~~~~~~~~~~~~~~~~~~~~~~~~~~~\nonumber\\
{1 \over z_{g}} {1 \over 3\pi^{3}}
\mu_{G}^{2}
\left|g_{S}(r_{1}^{(min)})
K_{1}(\mu_{G}\rho_{1}){\vec{\rho}_{1}\over \rho_{1}}
-g_{S}(r_{2}^{(min)})
K_{1}(\mu_{G}\rho_{2}){\vec{\rho}_{2} \over \rho_{2}}\right|^{2} \, .
\label{eq:6}
\endarr
Here $g_{S}(r)$ is the running colour charge,
$r_{1,2}^{(min)}={\rm min}\{r,\rho_{1,2}\}$, $K_{1}(x)$ is the
modified Bessel function,
$z_{g}$ is a fraction of the (lightcone) momentum of
$q\bar{q}$ pair carried by the gluon, and $\int dz_{g}/z_{g}
=log(s/s_{0})=\xi$.
With allowance for the $q\bar{q}g$ Fock state the
dipole cross section takes the form $\sigma_{tot}(\xi,r)=
\sigma_{0}(r)+\sigma_{1}(r)\xi$, where [6]
\arr
\sigma_{1}(r)=
\int d^{2}\vec{\rho}_{1} \,\,
z_{g}|\Phi_{1}(\vec{r},\vec{\rho}_{1},\vec{\rho}_{2},z_{g})|^{2}
\Delta\sigma_{g}(r,\rho_{1},\rho_{2})
 = {\cal K}\otimes \sigma_{0}(r)
\, \, .
\label{eq:7}
\endarr
To higher orders in $\xi$,
$
\sigma(\xi,r)=\sum_{n=0}{1\over n!}\sigma_{n}(r)
\xi^{n} $,
where $\sigma_{n+1}={\cal K}\otimes \sigma_{n}$, so that
\beq
{\partial \sigma(\xi,r) \over \partial \xi} ={\cal K}\otimes
\sigma(\xi,r)
\label{eq:8}
\endeq
is our generalized BFKL equation for the dipole
cross section. We emphasize that
the introduction of the gluon correlation length $R_{c}$
in the kernel ${\cal K}$
does not conflict the gauge-invariance constraints
$\sigma(r) \rightarrow 0$ and
$|\Phi(\vec{r}_{1},\vec{\rho}_{1},\vec{\rho}_{2},z_{g})|^{2}
\rightarrow 0$ at $r\rightarrow 0$, and
$\Delta\sigma_{g}(r,\rho_{1},\rho_{2}) \rightarrow 0$ at
$\rho_{1,2} \rightarrow 0$.
The essential ingredient of this derivation is
the subtraction of $\sigma_{0}(r)$ in Eq.~(\ref{eq:5}), which
in a simple and intuitively appealing form takes care of the
virtual radiative corrections.
Once the dipole-dipole scattering problem is solved, Eq.~(\ref{eq:4})
gives the hadron-hadron cross section.



In the BFKL scaling limit of $r, \rho_{1},\rho_{2} \ll R_{c}$
and fixed $\alpha_{S}$
\beq
\mu_{G}^{2}\left|K_{1}(\mu_{G}\rho_{1}){\vec{\rho}_{1}\over \rho_{1}}
-K_{1}(\mu_{G}\rho_{2}){\vec{\rho}_{2}\over\rho_{2}}\right|^{2} =
{r^{2} \over \rho_{1}^{2}\rho_{2}^{2}}   \, ,
\label{eq:9}
\endeq
the kernel ${\cal K}$ becomes independent of
$R_{c}$ and with the fixed $\alpha_{S}$ it takes on the
scale-invariant form. The BFKL
eigenfunctions of Eq.~(\ref{eq:8}) are
$
E(\omega,\xi,r)=(r^{2})^{{1\over 2}+\omega}
\exp[\xi\Delta(\omega)]
$
with the eigenvalue (intercept)
[here $\vec{r}=r\vec{n}$,  $\vec{\rho}_{1}=r\vec{x}$  and
$\vec{\rho}_{2}=r(\vec{x}+\vec{n})$]
\arr
\Delta(\omega) = {3\alpha_{S} \over 2\pi^{2}}
\int d^{2}\vec{x}~~{2(\vec{x}^{2})^{{1\over 2}+\omega}-1
\over \vec{x}^{2}
(\vec{x}+\vec{n})^{2}}=
{3\alpha_{S} \over \pi}
\int_{0}^{1} dz
{z^{{1\over 2}-\omega}+z^{{1\over 2}+\omega}-2z \over z(1-z)}
\nonumber\\
=
{3\alpha_{S} \over \pi} [2\Psi(1)-
\Psi({1\over 2}-\omega)-\Psi({1\over 2}+\omega)]\, , ~~~~~~~~~~~~~~
\label{eq:10}
\endarr
where $\Psi(x)$ is the digamma-function.
The final result for $\Delta(\omega)$ coincides with eigenvalues of
the BFKL equation. Indeed, in the scaling limit of
$\mu_{G}\rightarrow 0$ our Eq.~(\ref{eq:8}) can be transformed to
the same form as the original BFKL equation for the differential
distribution of gluons [1,2].

When $\omega$ is real and varies from $-{1\over 2}$ to
0 and to ${1\over 2}$, also the intercept
$\Delta(\omega)$ is real and varies from $+\infty$ down to
$\Delta(0)=\Delta_{\Pom}$ and back to $+\infty$, along the cut
from $j=1+\Delta_{\Pom}$ to $+\infty$ in
the complex angular momentum $j$ plane. If $\omega =i\nu$ and $\nu$
varies from $-\infty$ to
0 and to $+\infty$, then the intercept
$\Delta(i\nu)$ is again real and varies from $-\infty$ up to
$\Delta(0)=\Delta_{\Pom}$ and back to $-\infty$, along the cut
from $j=-\infty$ to $j=1+\Delta_{\Pom}$
in the complex $j$-plane. The choice of the latter cut
is appropriate for the Regge asymptotics at $\xi \gg 1$ and
the counterpart of the conventional Mellin representation is
\arr
\sigma(\xi,r)=
\int_{-\infty}^{+\infty} d\nu \,f(\nu)E(i\nu,r,\xi) = ~~~~~~~
\nonumber\\
r\int_{-\infty}^{+\infty} d\nu
f(\nu)\exp[2i\nu\log(r)]\exp(\Delta(i\nu)\xi)
\label{eq:11}
\endarr
where the spectral amplitude $f(\nu)$ is determined by the
boundary condition $\sigma(\xi=0,r)$:
\beq
f(\nu)={1\over \pi}\int dr{\sigma(0,r)\over r^{2}}
\exp[-2i\nu\log(r)] \,\,.
\label{eq:12}
\endeq
In the BFKL regime, the rightmost $j$-plane singularity
corresponds to the asymptotic cross section
\beq
\sigma_{\Pom}(\xi,r)  \propto r\exp(\xi\Delta_{\Pom}) \, .
\label{eq:13}
\endeq

The solution of Eq.~(\ref{eq:8}) can be written as
\beq
\sigma(\xi,r)=r\int {dr'\over (r')^{2}}K(\xi,r,r')\sigma(\xi=0,r')\, ,
\label{eq:14}
\endeq
where in the BFKL regime the evolution kernel equals
\arr
K(\xi,r,r')={1\over \pi}\int d\nu \exp\left[2i\nu\log{r\over r'}
\right]\exp[\xi \Delta(i\nu)]
\nonumber\\
\propto {\exp(\Delta_{\Pom}\xi) \over \sqrt{\xi}}
\exp\left(-2{(\log r-\log r')^{2}\over
\xi \Delta^{''}(0)}\right) \, .
\label{eq:15}
\endarr
The `diffusion' kernel (\ref{eq:15}) makes it obvious that
starting with $\sigma(\xi=0,r)$ which was concentrated at the
perturbative small $r \lsim R \ll R_{c}$ one ends up at large
$\xi$ with $\sigma(\xi,r)$ which extends up to the
nonperturbative $r\sim R\exp(\sqrt{\xi\Delta''(0)}) > R_{c}$ .
This `diffusion' towards large $r$ is
further accelerated if the
running coupling is introduced.
Thus, the scattering of even very small dipoles of size
$r\sim R \ll R_{c}$ and/or deep inelastic scattering at
$Q^{2}\gg R_{c}^{-2}$, will eventually be dominated by interactions
of the perturbative gluons sticking out of small dipoles at a
distance $\rho \sim R_{c}$, and $\Delta_{\Pom}$ will be determined
by interactions at the scale $R_{c}$ and by the frozen coupling
$\alpha_{S}^{(fr)}$.



The analytic solution of Eq.~(8) at finite $R_{c}$ and with the
running coupling is not available, and we resort to the numerical
analysis. We use the running QCD coupling
$
\alpha_{S}(r)=6\pi/[(33-2N_{f})\log(
1/\Lambda r)]
$
with $\Lambda=0.2 GeV$,
and at large $r$ we impose the simplest freezing
$
\alpha_{S}(r)=\alpha_{S}^{(fr)}=
{\rm min}\{\alpha_{S}(R_{c}),1\} \, .
$
Firstly, we calculate $\Delta_{eff}(r,R)=
\sigma_{1}(r,R)/\sigma_{0}(r,R)$ for the boundary condition
$\sigma(\xi=0,r)$ given by the dipole-dipole cross section
(\ref{eq:2}). The first interesting case
is the scattering of the equal-size dipoles (Fig.~1).
Here the kernel ${\cal K}$ suggests the
quasiclassical estimate for the effective intercept
\beq
\Delta_{\Pom}(r) ={12\log2\over \pi}\alpha_{S}(r)\,.
\label{eq:16}
\endeq
Indeed, we find that at small $r$ the ratio
$\beta=\Delta_{eff}(r,r)/\Delta_{\Pom}(r)$ tends to a constant
$\beta \approx 0.57$ independent of the gluon correlation radius
$R_{c}$.
The effective intercept $\Delta_{eff}(r,r)$ rises with $r$ up
to $r \sim R_{c}$, then decreases and flattens at $r\gg R_{c}$,
where both $\sigma_{0,1}(r) \propto R_{c}^{2}$.

Another interesting regime is the DLLA of unequal dipoles
$r \ll R$. In this limit Eq.~(\ref{eq:7}) takes the form [6]
\beq
\sigma_{n+1}(r)={\cal K}\otimes \sigma_{n}(r)=
{3r^{2}\alpha_{S}(r) \over \pi^{2}}\int_{r^{2}}^{R_{c}^{2}}
{d^{2}\vec{\rho} \over \rho^{4}}\sigma_{n}(\rho)\, ,
\label{eq:17}
\endeq
which is equivalent to the GLDAP evolution equation [8] and
to the first order in $\xi$ gives
$\Delta_{DLLA}(r,R) = {2\over 3}\log[\alpha_{S}(R)/\alpha_{S}(r)]$.
Like any logarithmic estimate, this formula works up to a
constant term $\sim 1$. In Fig.~2 we show our results for
$\Delta_{eff}(r,R)$ for $R=1{\rm f}$. The difference
$\delta=\Delta_{eff}(r,R)-\Delta_{DLLA}(r,R)$ flattens
at $r/R\lsim 0.2$, which is a signal of the onset of DLLA.

The position of the rightmost singularity in the $j$-plane
can easily be estimated from the asymptotic behavior of the
numerical solution of Eq.~(\ref{eq:7}). At $R_{c}=0.4,~0.28,~0.22~$f,
i.e., at the frozen coupling $\alpha_{S}^{(fr)}=1.0,~82,~0.63$,
we find estimates $\Delta_{\Pom}\approx 0.52,~0.41,~0.36$,
respectively.
The detailed discussion of the convergence to,
and properties of, the limiting cross section $\sigma_{\Pom}(\xi,r)$
will be presented elsewhere. We only notice, that these estimates
are significantly below the Collins-Kwiecinski lower bound
$\Delta_{\Pom} > 3.6\alpha_{S}^{(fr)}/\pi$ [9]. (The derivation
of this bound in [9] is flawed by the infrared cutoff which
breaks the initial symmetry of the BFKL kernel in the momentum
space.)

The two cases of certain theoretical, although of little practical,
interest are worth of mention. The first is the case of finite
$R_{c}$ at fixed $\alpha_{S}$, the second case is of
$\mu_{G} \rightarrow 0$ in the wave function (\ref{eq:6}) while
keeping finite $R_{c}$ in the strong coupling. Both share the
property of restoration of the scaling invariance of the kernel
${\cal K}$ on the infinite semiaxis
$\log r < \log R_{c}$ or $\log r > \log R_{c}$,
where $\alpha_{S}(r)$ freezes, respectively.
On the corresponding semiaxis, the
eigenfunctions are essentially identical to the BFKL set,
the spectrum of eigenvalues will evidently
be continuous and the $j$-plane partial waves will have the
cut in the $j$-plane. Here we differ
from Lipatov [3], who concluded that the running coupling leads
to the discret spectrum of eigenvalues
and to a sequence of poles in the $j$-plane. The numerical analysis
shows that the tip of the cut in the $j$-plane is very close
to $\Delta_{\Pom}$ as given by Eq.~(\ref{eq:1})
with $\alpha_{S}=\alpha_{S}^{(fr)}$; the more detailed analysis is
needed to check a possibility of a finite departure from
Eq.~(\ref{eq:1}) which depends on the value of $\alpha_{S}^{(fr)}$.

Finally, let us consider the $\pi N$ interaction as the typical
hadronic scattering.
The plausible assumption is that the growth of the
hadronic cross sections is dominated by the perturbative gluons.
In this case, $\sigma_{1}^{(pt)}(\pi N)=\langle \langle
\sigma_{1}(r_{\pi},r_{N}) \rangle_{N}\rangle_{\pi}$ and
$
\Delta_{\Pom}(\pi N) \approx \sigma_{1}^{(pt)}(\pi N)/
\sigma_{tot}(\pi N)$.  In Fig.3 we present our
prediction for the perturbative QCD contribution to the total
cross section $\sigma_{0}(\pi N)$ and the
$\Delta_{eff}(\pi N)$ vs. the gluon correlation radius
$R_{c}$.
We reproduce the empirical value $\Delta_{\Pom}(hN) \sim 0.1$
at $R_{c}\sim 0.4$fm, when  $\sim 40\%$ of
$\sigma_{tot}(\pi N)$ is of the perturbative origin.

Besides the rise of the total cross section, variations of the
dipole cross section for the presence of gluons also contribute
to the triple-pomeron coupling $A_{3\Pom}$. The method for
calculating $A_{3\Pom}$ was presented in [6,7].
With a correlation length $R_{c}\sim 0.4 $fm,
we find $A_{3\Pom}(\pi N)\sim 0.04{\rm GeV}^{-2}$,
which is conistent with the experimental determinations [10].

In conclusion, we have derived generalized BFKL
equation for the total cross sections with allowance for the
finite gluon correlation radius $R_{c}$. We presented the first
estimates of the perturbative $QCD$ contribution to the
rate of the growth of $\sigma_{tot}(pN)$ and to the triple-pomeron
coupling, which are consistent with experiment if
$R_{c}\sim 0.4$ fm. (Incidentally, the instanton model of the QCD
vacuum and the lattice QCD calculations give
very close value of $R_{c}$ [11].) The
irrefutable advantage of having the
equation for the total cross section
and of using the dipole-size representation
which diagonalizes the scattering matrix, is that they allow an easy
incorporation of the unitarity constraints. To this end, we
recall that consideration of the unitarity corrections in the
DLLA limit has already lead to an important conclusion [6] that
the unitarity correction to structure functions in the diffractive
deep inelastic scattering at small $x$ satisfies the linear GLDAP
evolution equations.

{\bf Acknowledgements}: We are grateful to L.N.Lipatov for
numerous discussions on the pomeron, and to S.Bass for useful
suggestions. B.G.Z. and V.R.Z. are grateful to J.Speth for the
hospitality at IKP, KFA J\"ulich.
\pagebreak

\pagebreak
\begin{itemize}
\item[Fig.1 - ]
The effective intercept $\Delta_{eff}(r,r)$ for the scattering of
two identical dipoles of size $r$ in comparison with
$\Delta_{\Pom}(r)$ Eq.~(16). The lower box shows
the ratio $\Delta_{eff}(r,r)/\Delta_{\Pom}(r)$.
The curves
{\sl a), b)} and {\sl c)} are for $\mu_{G}=0.3,~0.5,~0.7~$GeV,
respectively.

\item[Fig.2 - ]
The effective intercept $\Delta_{eff}(r,R)$ for the scattering of
unequal dipoles of size $r$ and $R$ in comparison with DLLA
formula $\Delta_{DLLA}(r,R)$. The curves
{\sl a),~b)} are for $\mu_{G}=0.3,~0.5$GeV,
respectively.
\item[Fig.3 - ]
The effective intercept $\Delta_{\Pom}(\pi N)$, the perturbative
QCD contribution to the total cross section
$\sigma_{0}(\pi N)$ and the effective triple-pomeron
coupling for the pion-nucleon scattering vs. $\mu_{G}$.
\end{itemize}
\end{document}